\documentclass[%
reprint,
preprintnumbers,
nofootinbib,
amsmath,amssymb,longbibliography,
aps
prd
]{revtex4-1}
\pdfoutput=1
\usepackage{graphicx}
\usepackage[utf8]{inputenc}
\usepackage{flushend}
\usepackage{dcolumn}
\usepackage{bm}
\usepackage{soul}
\usepackage{balance}
\usepackage[normalem]{ulem}
\usepackage[colorlinks = true,
            linkcolor = blue,
            urlcolor  = blue,
            citecolor = blue,
            anchorcolor = blue]{hyperref}
\usepackage{verbatim,subfigure}
\usepackage{color,ulem}
\usepackage[english]{babel}
\usepackage{MnSymbol,wasysym}
\usepackage[utf8]{inputenc}
\input Starburst.fd
\newcommand*\initfamily{\usefont{U}{Starburst}{xl}{n}}\initfamily 

\newcommand{\beq}{\begin{eqnarray}}
\newcommand{\eeq}{\end{eqnarray}}
\usepackage{amsmath}
\usepackage{tikz}
\usetikzlibrary{decorations.pathmorphing}
\usetikzlibrary{shapes.misc}
\tikzset{cross/.style={cross out, draw=black, minimum size=8*(#1-\pgflinewidth), inner sep=0pt, outer sep=0pt},
cross/.default={1pt}}
\usetikzlibrary{patterns,math}
\definecolor{applegreen}{rgb}{0.55, 0.71, 0.0}
\usepackage{contour}
\usepackage{xparse}
\ExplSyntaxOn
\NewDocumentCommand{\HS}{m}
 {
  \seq_set_split:Nnn \l_tmpa_seq { ~ } { #1 }
  \seq_map_inline:Nn \l_tmpa_seq { \contour{green}{##1} ~ } \unskip
 }
\ExplSyntaxOff

\definecolor{darkviolet}{rgb}{0.58, 0.0, 0.83}
\definecolor{mygreen}{rgb}{0.0, 0.5, 0.0}

\begin{document}
\preprint{\texttt{\normalsize{IFT-UAM/CSIC-25-23}}}

%
\title{Singular Value Decomposition and its Blind Spot for Quantum Chaos\\
in Non-Hermitian Sachdev-Ye-Kitaev Models}

\author{\vspace{-1.6mm}Matteo Baggioli,$^{1,2,3}$ Kyoung-Bum Huh,$^{2}$ Hyun-Sik Jeong,$^{4}$ Xuhao Jiang,$^{2,5,6}$ Keun-Young Kim$^{7,8}$ and Juan F. Pedraza$^{4}$}

\affiliation{\vspace{4pt}$^{1}$School of Physics and Astronomy, Shanghai Jiao Tong University, Shanghai 200240, China}
\affiliation{$^{2}$Wilczek Quantum Center, School of Physics and Astronomy, Shanghai Jiao Tong University, Shanghai 200240, China}
\affiliation{$^{3}$Shanghai Research Center for Quantum Sciences, Shanghai 201315, China}
\affiliation{$^{4}$Instituto de F\'isica Te\'orica UAM/CSIC, Calle Nicol\'as Cabrera 13-15, 28049 Madrid, Spain}
\affiliation{$^{5}$Technical University of Munich, TUM School of Natural Sciences, Physics Department, 85748 Garching, Germany}
\affiliation{$^{6}$Faculty for Physik, Ludwig-Maximilians-Universit\"at M\"unchen, 
Schellingstraße 4, 80799, Munich, Germany}
\affiliation{$^{7}$Department of Physics and Photon Science, Gwangju Institute of Science and Technology, Gwangju 61005, Korea}
\affiliation{$^{8}$Research Center for Photon Science Technology, Gwangju Institute of Science and Technology, Gwangju 61005, Korea\vspace{-1.6mm}}

\begin{abstract}
The study of chaos and complexity in non-Hermitian quantum systems poses significant challenges due to the emergence of complex eigenvalues in their spectra. Recently, the singular value decomposition (SVD) method was proposed to address these challenges. In this work, we identify two critical shortcomings of the SVD approach when analyzing Krylov complexity and spectral statistics in non-Hermitian settings. First, we show that SVD fails to reproduce conventional eigenvalue statistics in the Hermitian limit for systems with non-positive definite spectra, as exemplified by a variant of the Sachdev-Ye-Kitaev (SYK) model. Second, and more fundamentally, Krylov complexity and spectral statistics derived via SVD cannot distinguish chaotic from integrable non-Hermitian dynamics, leading to results that conflict with complex spacing ratio analysis. Our findings reveal that SVD is inadequate for probing quantum chaos in non-Hermitian systems, and we advocate employing more robust methods, such as the bi-Lanczos algorithm, for future research in this direction.
\end{abstract}

\maketitle
%
\textbf{Introduction.}
In recent years, the study of open quantum systems~\cite{Rivas_2012,Fazio:2024aa} has attracted considerable attention due to the novel physical phenomena arising from interactions with the environment. Such interactions inherently render these systems non-Hermitian, complicating their theoretical description and necessitating methods beyond those developed for closed systems. Unlike standard quantum systems, non-Hermitian Hamiltonians, particularly those with broken parity-time (PT) symmetry, feature complex eigenvalues, posing the fundamental challenge of interpreting their physical significance~\cite{Bender_2007}.

Characterizing quantum chaos in closed systems fundamentally relies on analyzing their spectra. Due to the emergence of complex eigenvalues in non-Hermitian open systems, conventional diagnostics for quantum chaos cannot be directly applied. In closed systems, for example, spectral statistics distinguish integrability from chaos using Poisson and Wigner distributions, respectively. This distinction is formalized by the Bohigas-Giannoni-Schmit conjecture~\cite{Bohigas:1983er}, where level repulsion is governed by the Dyson index. For non-Hermitian systems, several extensions that incorporate complex eigenvalues have been proposed~\cite{Grobe:1988zz,Hamazaki:2020kbp,Sa:2020fpf}. Chaotic open quantum systems, in particular, are effectively described by Ginibre statistics~\cite{Ginibre:1965zz}, exhibiting universal cubic level repulsion.

However, the analysis of complex eigenvalues presents technical difficulties~\cite{Akemann:2019aa}, particularly in the necessary processes of unfolding spectra and defining meaningful level statistics. To overcome these issues, alternative spectral observables that avoid unfolding, such as complex spacing ratios (CSR)~\cite{Sa:2020fpf} and dissipative spectral form factors~\cite{Fyodorov:1997aa,Li:2021kuv,Li:2024aa}, have been developed.

An alternative strategy to circumvent the problem of directly analyzing complex eigenvalues is to use singular values ($\sigma$) obtained through singular value decomposition (SVD) \cite{Kawabata:2023aa,Roccati:2023aa,Hamanaka:2024aa,Tekur:2024aa,Nandy:2024aa,Nandy:2025aa}. Singular values, being real and non-negative by construction, allow for a straightforward extension of Hermitian spectral diagnostics into non-Hermitian settings via one-dimensional statistical descriptions. This approach has been employed to study singular spacing ratios $\langle r_\sigma \rangle$ \cite{Kawabata:2023aa}, singular spectral form factors (SFF$_\sigma$) \cite{Roccati:2023aa,Nandy:2024aa}, and various measures of spectral and Krylov complexity \cite{Nandy:2024aa,Nandy:2025aa}. Crucially, it has been argued that both singular value statistics and the associated Krylov complexity serve as effective tools for probing quantum chaos and capturing integrable-to-chaotic transitions in non-Hermitian quantum models.

In this work, we investigate the prototypical non-Hermitian $q$-body Sachdev-Ye-Kitaev (nHSYK) model~\cite{Garcia-Garcia:2020ttf,Su:2020quk,Liu:2020fbd,Garcia-Garcia:2021elz,Zhang:2021klq,Garcia-Garcia:2021rle,Garcia-Garcia:2022xsh,Cipolloni:2022fej}, which plays a central role in holographic studies. We identify significant inconsistencies arising when employing the SVD approach to compute spectral statistics and Krylov complexity. Specifically, the SVD method fails both to reliably probe quantum chaos and to correctly reproduce the Hermitian limit. Our results thus call into question the physical significance and general validity of the SVD-based observables.

%
\vspace{1mm}
\textbf{Non-Hermitian SYK model.}
We consider the non-Hermitian SYK (nHSYK) model with $N$ Majorana fermions and $q$-body interactions, introduced in~\cite{Garcia-Garcia:2021rle,Garcia-Garcia:2022xsh}:
\begin{align}\label{SYKMODEL}
    H = \sum_{i_1<i_2< \cdots < i_q }^{N}\, (J_{i_1\,i_2\,\cdots\,i_q}+i M_{i_1\,i_2\,\cdots\,i_q})\, \psi_{i_1}\, \psi_{i_2}\, \cdots \psi_{i_q} \,,
\end{align}
where $\psi_i$ are Majorana fermions satisfying $\{\psi_i, \psi_j\} = \delta_{ij}$. The couplings $J$ and $M$ are Gaussian-distributed random variables with zero mean and variance $\langle J^2\rangle = \langle M^2\rangle = (q-1)!/N^{q-1}$. A nonzero coupling $M$ explicitly breaks Hermiticity. In this manuscript, we consider $N=26$ for $q=4$ and $q=2$, incorporating a parity-symmetry block that reduces the effective dimension to $d=2^{{N/2}-1}$.

In Hermitian quantum systems, spectral statistics distinguish between integrable and chaotic regimes. Integrable models typically exhibit one-dimensional Poisson statistics, with level-spacing distribution given by $p(s) = e^{-s}$~\cite{BerryTabor}. In contrast, chaotic systems follow Wigner statistics, characterized by random matrix theory (RMT) distributions exhibiting level repulsion of the form $p(s)\approx s^\beta$ as $s \to 0$. Here, $\beta = 1,\,2,\,4$ correspond to the Gaussian Orthogonal Ensemble (GOE), Gaussian Unitary Ensemble (GUE), and Gaussian Symplectic Ensemble (GSE)~\cite{Bohigas:1983er, Meh2004}, respectively.
\begin{figure}[t!]
 \centering
 \vspace{1.2mm}
     {\includegraphics[width=4.2cm]{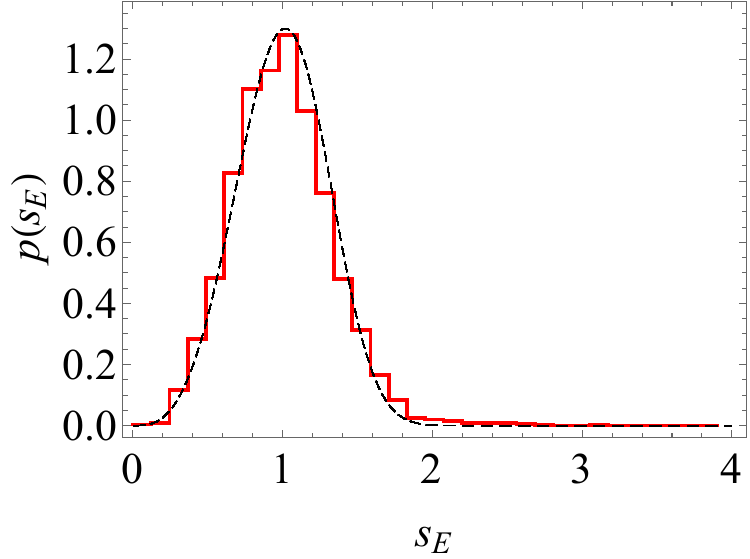}}
     {\includegraphics[width=4.2cm]{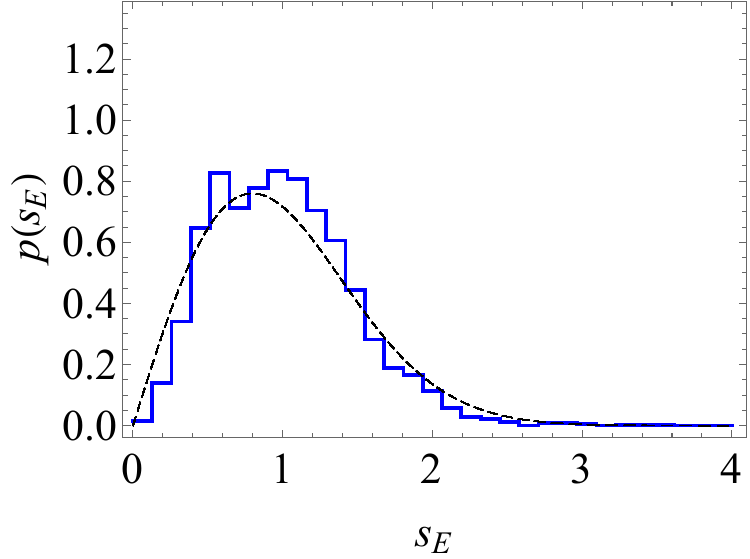}}
\caption{Complex eigenvalue spacing distributions for the nHSYK model with $q=4$ (left) and $q=2$ (right). The $q=4$ case follows GinUE statistics~\eqref{GinUEdis}, indicative of quantum chaos, whereas the $q=2$ case exhibits two-dimensional Poisson statistics~\eqref{2DPoidis}, consistent with integrable dynamics.}\label{complexLSDfig}
\end{figure}

This classification can be generalized to non-Hermitian open quantum systems~\cite{Grobe:1988zz}. Integrable cases in these systems follow a two-dimensional Poisson distribution for the nearest-neighbor spacing of complex eigenvalues:
\begin{align}\label{2DPoidis}
   p(s) = \frac{\pi}{2} \, s \, e^{-\frac{\pi}{4}s^2}\,.
\end{align}
Unlike the Hermitian case, the linear repulsion $p(s)\approx s$ as $s \to 0$ is not due to eigenvalue correlations but instead simply reflects the finite area occupied by eigenvalues. In contrast, chaotic non-Hermitian systems typically follow Ginibre statistics~\cite{Ginibre:1965zz,Hamazaki:2020kbp}, described by:
\begin{align}\label{GinUEdis}
   p(s) = \left( \prod_{k=1}^{\infty} \frac{\Gamma\left(1+k,s^2\right)}{k!}  \right) \sum_{j=1}^{\infty} \frac{2 s^{2j+1} e^{-s^2}}{\Gamma\left(1+j,s^2\right)} \,,
\end{align}
where $\Gamma\left(1+k,s^2\right) = \int_{s^2}^{\infty} t^k e^{-t} d t$ is the incomplete gamma function. Notably, unlike Hermitian RMT, all three Ginibre ensembles (GinUE, GinOE and GinSE) share the same spacing distribution~(\ref{GinUEdis}), exhibiting cubic repulsion, $p(s) \approx s^3$ as $s \to 0$ \cite{Grobe:1988zz,Grobe:1989aa,Akemann:2019aa}. More generally, Ginibre statistics define one of three universality classes (class A) in non-Hermitian RMT, whereas the other two classes, AI$^{\dagger}$ and AII$^{\dagger}$, are distinguished by their transposition symmetries. With our chosen parameters, the nHSYK model belongs to class A (GinUE). For a more in-depth analysis covering various values of $N$ and $q$, see~\cite{Garcia-Garcia:2021rle}.

We have analyzed the complex eigenvalue spacing distributions for our nHSYK model, with the results presented in Fig.~\ref{complexLSDfig}. Upon performing a naive unfolding of the complex spectra, we observe that for $q=4$, the spacing distribution closely matches GinUE statistics \eqref{GinUEdis}, signaling quantum chaotic behavior. In contrast, the distribution for $q=2$ aligns with the two-dimensional Poisson statistics~\eqref{2DPoidis}, indicative of integrability. 

\begin{figure}[t!]
 \centering
     {\includegraphics[width=4.0cm]{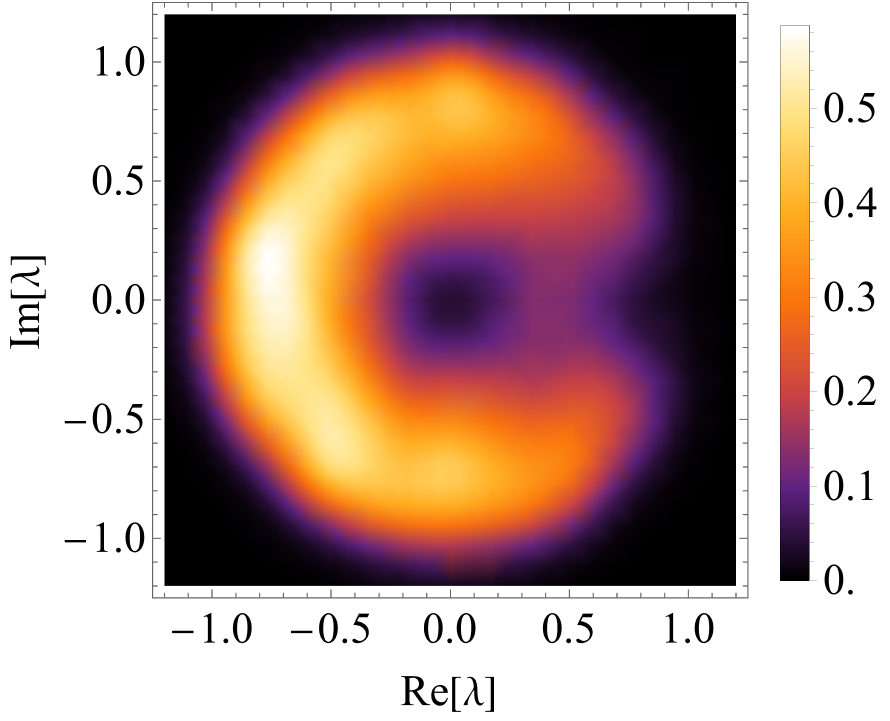}}\hspace{3mm}
     {\includegraphics[width=4.0cm]{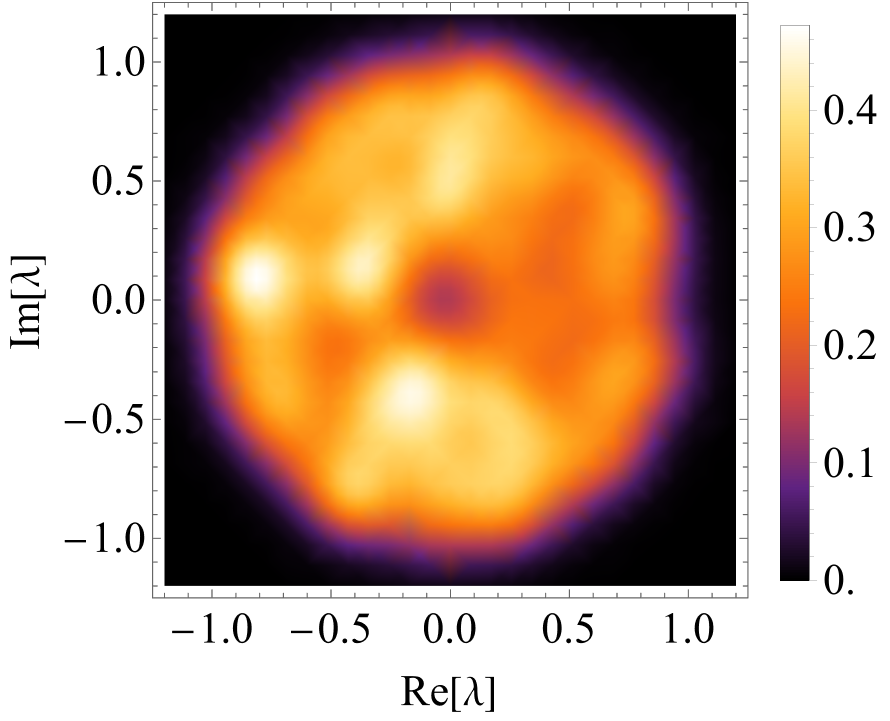}}
\caption{Distribution of complex spacing ratios for the nHSYK model with $q=4$ (left) and $q=2$ (right). The $q=4$ case exhibits level repulsion, characteristic of quantum chaos, while the $q=2$ case shows a nearly uniform distribution, indicative of integrable behavior.}\label{complexSRfig}
\end{figure}

In addition, we examined the complex spacing ratio (CSR)~\cite{Sa:2020fpf}, a non-Hermitian extension of the conventional $\langle r \rangle$-value~\cite{Atas:2013aa}, defined as
\begin{align}\label{CSRfor}
   \lambda_{k} = \frac{E^{\text{NN}}_k - E_k}{E^{\text{NNN}}_k - E_k} \,,
\end{align}
where $E_k$ are complex eigenvalues, and $E^{\text{NN}}_k$ and $E^{\text{NNN}}_k$ represent the nearest and next-to-nearest neighbors, respectively. The CSR method eliminates the necessity of unfolding and, by construction, satisfies $|\lambda_{k}|\leq1$, ensuring all complex spacing ratios lie within the unit disk.
In Fig.~\ref{complexSRfig}, we present the distribution function of the CSR, Eq.~\eqref{CSRfor}, for the nHSYK model with $q=4$ and $q=2$.
For $q=4$, the CSR shows strong suppression at small spacings, particularly at small angles, indicating level repulsion characteristic of quantum chaos. In contrast, for $q=2$, the distribution is nearly uniform, consistent with integrability. These findings align with non-Hermitian RMT predictions~\cite{Sa:2020fpf}: specifically, in chaotic systems, the CSR follows a radial distribution $\rho(r) \approx r^3$ (cubic level repulsion) and exhibits anisotropy in the angle distribution $\rho(\theta)$, quantified by $\langle \cos \theta \rangle = \int d \theta \, \cos \theta \,\rho(\theta) \approx 0.24$. In contrast, uncorrelated levels in integrable cases follow $\rho(r) \approx r$ with $\langle \cos \theta \rangle =0$. Our results confirm these predictions in the nHSYK model, with $q=4$ yielding $\langle \cos \theta \rangle \approx 0.25$ and $q=2$ giving $\langle \cos \theta \rangle \approx 0.07$.

Thus, well-established complex energy level statistics confirm that our model is chaotic for $q=4$ and integrable for $q=2$. Notably, at least in this model, non-Hermitian deformations do not seem to affect the chaotic or integrable nature of the corresponding Hermitian system.

%
\textbf{Singular value decomposition.}
Singular value decomposition (SVD) expresses a generic non-Hermitian Hamiltonian as $H = U \Sigma V^{\dagger}$, where $U$ and $V^{\dagger}$ are unitary matrices formed by the orthonormal eigenvectors of $H H^{\dagger}$ and $H^{\dagger} H$, respectively. The diagonal matrix $\Sigma=\text{diag} \left(\sigma_1,\, \cdots,\, \sigma_d \right)$ contains the singular values $\sigma_i$, which are real and non-negative~\cite{Nielsen_2012}. In Hermitian systems, singular values correspond to the absolute values of the eigenvalues, whereas in non-Hermitian systems the relationship between the two is more intricate~\cite{Allard:2024aa}.
Singular values can alternatively be obtained via the Hermitization method~\cite{FEINBERG1997579}, by constructing the effective Hamiltonian $H_{\text{eff}}\equiv\sqrt{H^{\dagger} H}$ or $\sqrt{H H^{\dagger}}$ so that its eigenvalues represent directly the singular values of $H$.

Recently, Ref.~\cite{Kawabata:2023aa} proposed using singular value statistics instead of complex eigenvalues in non-Hermitian systems, allowing the application of standard quantum chaos tools from Hermitian systems. Singular value spacing distributions, $p(s_\sigma)$, are expected to follow Poisson statistics in integrable systems and, for instance, GUE distributions in chaotic ones, respectively:
\begin{align}\label{sigmaStatistics}
   p(s_\sigma) = e^{-s_\sigma}\,, \qquad 
   p(s_\sigma) = \frac{32}{\pi^2} \, s_\sigma^2 \, e^{-\frac{4}{\pi}s_\sigma^2}\,.
\end{align}
Other key measures include the singular spacing ratio,
\begin{align}\label{}
   \langle r_\sigma \rangle = \text{Mean} \left[ \frac{\text{min}\left(s_n,\,s_{n+1}\right)}{\text{max}\left(s_n,\,s_{n+1}\right)} \right] \,, \quad s_n = \sigma_{n+1} - \sigma_{n} \,,
\end{align}
and the singular spectral form factor, defined as
\begin{align}\label{sigmaSFF}
\begin{split}
\text{SFF}_\sigma = \frac{|Z(t)|^2}{|Z(0)|^2}\,, \qquad Z(t) = \text{Tr} \left[e^{i t H_{\text{eff}}}\right] \,.
\end{split}
\end{align}

\begin{figure}[t!]
 \centering
     {\includegraphics[width=4.2cm]{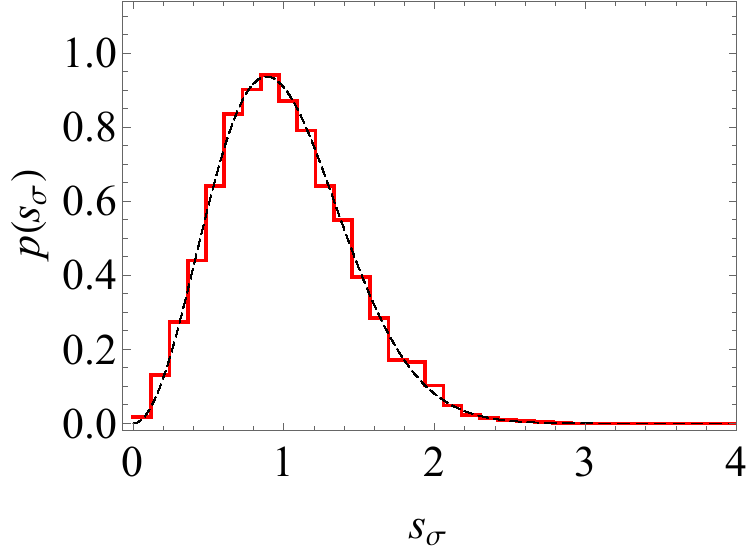}}
     {\includegraphics[width=4.2cm]{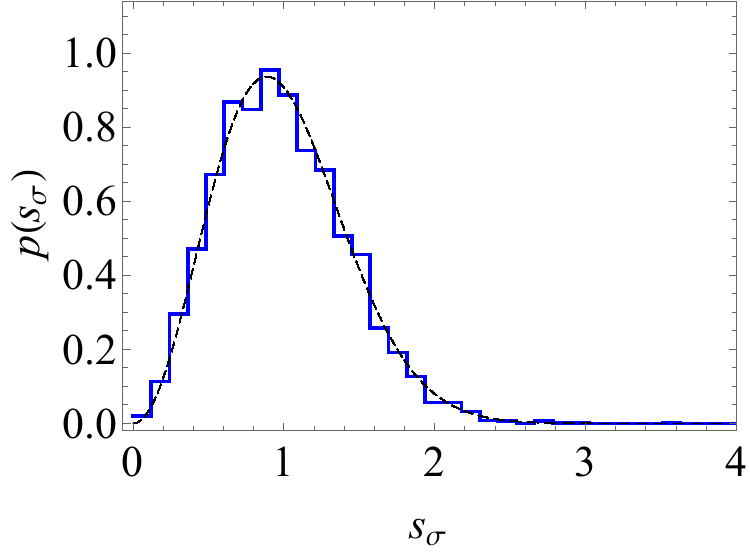}}
     
     {\includegraphics[width=4.2cm]{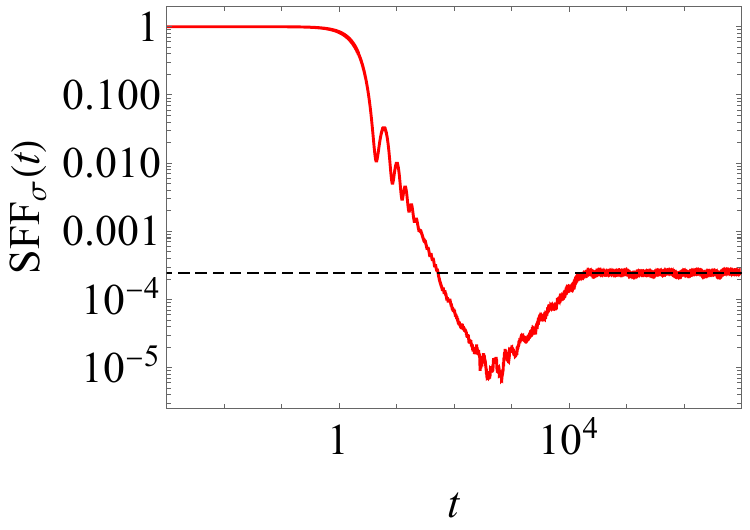}}
     {\includegraphics[width=4.2cm]{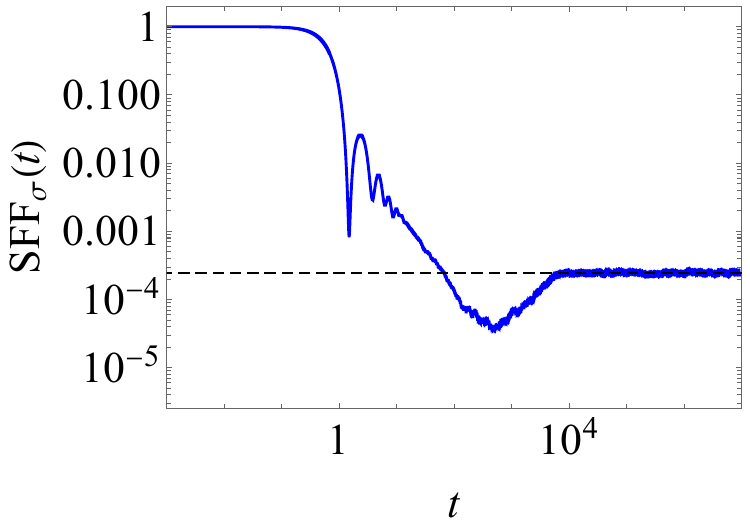}}   
\caption{The top panels show the singular level spacing distribution for $q=4$ (red) and $q=2$ (blue), while the bottom panels display the singular spectral form factor. The dashed lines in SFF$_{\sigma}$ represent the plateau value $1/d$. To smooth out short-time fluctuations and extract the long-term trend, we applied a simple moving average to SFF$_{\sigma}$.}
\label{SingularStafig}
\end{figure}

Using these concepts, we have analyzed singular value statistics in our non-Hermitian SYK model. The top panels of Fig.~\ref{SingularStafig} show the singular level spacing distribution, revealing quantum chaos signatures in both the $q=4$ and $q=2$ cases, which follow the GUE distribution. Additionally, both cases exhibit $\langle r_\sigma \rangle \approx 0.6$, consistent with RMT predictions, and a linear ramp in SFF$_\sigma$, as shown in the bottom panels of Fig.~\ref{SingularStafig}. The plateau of SFF$_\sigma$, $1/d$, agrees with the expected value in the Hermitian theory~\cite{Cotler:2016fpe}.

Our findings highlight a subtle relationship between complex energy eigenvalues and singular values in non-Hermitian systems~\cite{Allard:2024aa}. In fact, complex eigenvalue spacing distributions can distinguish the chaotic case, $q=4$ (GinUE), from the integrable case, $q=2$ (two-dimensional Poisson). On the other hand, singular level spacing distributions, as well as the singular spectral form factor, fail in this regard, as they cannot differentiate between the two cases and, therefore, cannot effectively probe quantum chaos.
These results reveal critical limitations of the singular value decomposition method for studying quantum chaos in non-Hermitian quantum systems. As we will see below, the Krylov state complexity built on singular values also suffers from the same issue. Nevertheless, it is interesting to note the internal consistency within singular statistics~\cite{Nandy:2024aa,Nandy:2025aa}: when the level spacing distribution exhibits apparent signatures of quantum chaos, $\langle r_\sigma \rangle$ and SFF$_\sigma$ do so as well.

%
\vspace{1mm}
\textbf{Krylov complexity of states based on SVD.}
Krylov complexity has emerged as a powerful tool for diagnosing quantum chaos~\cite{Balasubramanian:2022tpr}. It has been extensively studied in various quantum chaotic models, including RMT~\cite{Balasubramanian:2022tpr,Balasubramanian:2023kwd,Caputa:2024vrn,Bhattacharjee:2024yxj,Jeong:2024oao} and different versions of the SYK model~\cite{Erdmenger:2023wjg,Chapman:2024pdw,Baggioli:2024wbz,Huh:2024lcm}. For time-evolved thermofield double (TFD) states in chaotic systems, Krylov complexity typically exhibits four distinct phases: initial growth, a peak, a decline, and a plateau~\cite{Balasubramanian:2022tpr}. Notably, the peak is a hallmark of chaotic many-body systems~\cite{Baggioli:2024wbz} and, although generally absent in integrable models, can sometimes arise from certain types of instabilities~\cite{Huh:2023jxt}.

Our goal is to determine whether singular value-based Krylov complexity can similarly distinguish between chaos and integrability in nHSYK models, focusing on $q=4$ and $q=2$. Preliminary work in this direction~\cite{Nandy:2025aa} has shown that this complexity exhibits a peak for $q=4$ nHSYK, suggesting it may retain sensitivity to chaotic dynamics. A full characterization thus requires extending these results to the nonchaotic case, $q=2$.

To define singular value-based Krylov complexity, we construct the Krylov basis $\{|K_n \rangle\}$ using the Lanczos algorithm~\cite{Lanczos:1950zz,RecursionBook}, which generates the Lanczos coefficients $\{a_n,b_n\}$ that encode the system's dynamics. In the Hermitian case, these coefficients define the tridiagonal matrix representation of the Hamiltonian,
\begin{align}\label{}
     H_{\text{Hermitian}} |K_n \rangle = a_n | K_n \rangle + b_{n+1} | K_{n+1} \rangle + b_n | K_{n-1} \rangle \,.
\end{align}
In the non-Hermitian case, we consider an effective Hamiltonian $H_{\text{eff}}$ constructed from the non-Hermitian Hamiltonian $H$:
\begin{align}\label{}
\begin{split}
H_{\text{eff}} = \sqrt{H^{\dagger} H} \,,
\end{split}
\end{align}
which plays the role of $H_{\text{Hermitian}}$.
The Krylov wavefunctions $\Psi_n(t)$ then evolve via the Schrödinger equation,
\begin{align}\label{DES}
    i \, \partial_t \Psi_n(t) = a_n \Psi_n(t) + b_{n+1} \Psi_{n+1}(t) + b_n \Psi_{n-1}(t) \,,
\end{align}
yielding the time-evolved state $|\Psi(t) \rangle = \sum_n \Psi_n(t) | K_n \rangle$.

The `SVD Krylov complexity' is then defined as
\begin{align}\label{eq:Krylov complexity}
    C_{\sigma}(t)\equiv \sum_{n} n \, |\Psi_n(t)|^2 \,,
\end{align}
and quantifies the spread of the wavefunction in Krylov space.
Following Ref.~\cite{Balasubramanian:2022tpr}, we take the TFD state as the initial state, defined in terms of singular values.

\begin{figure}[t!]
 \centering
     {\includegraphics[width=4.2cm]{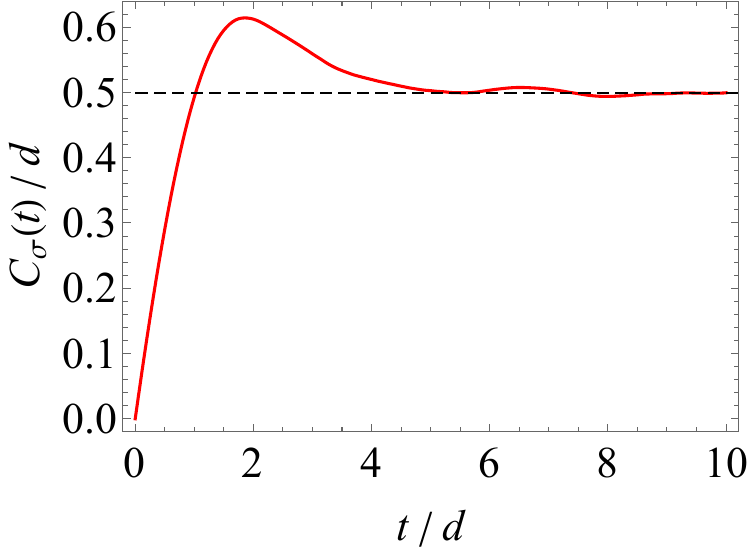}}
     {\includegraphics[width=4.2cm]{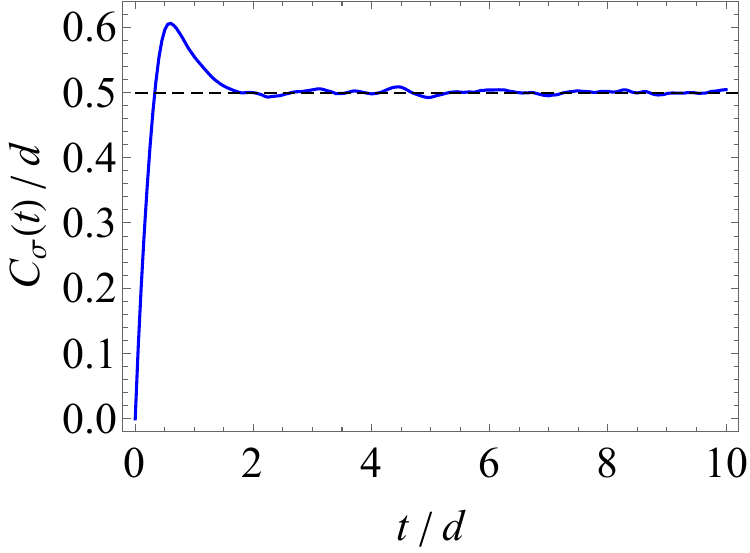}}
\caption{SVD Krylov complexity for $q=4$ (left) and $q=2$ (right). Both figures exhibit a characteristic peak typical of chaotic systems, highlighting that SVD Krylov complexity fails to distinguish between integrability and chaos. The dashed lines represent the saturation value, $C_\sigma(t=\infty) = d/2$.} \label{SingularKrylovfig}
\end{figure}

Figure~\ref{SingularKrylovfig} presents the SVD Krylov complexity for $q=4$ and $q=2$ in nHSYK models. In both cases, we observe a characteristic peak, suggesting a misleading inference of quantum chaos, yet consistent with singular value statistics. This peak occurs at a timescale $t_{\text{peak}}\approx t_H \approx \mathcal{O}(d)$, aligning with Hermitian systems~\cite{Camargo:2024deu}, where $t_H$ denotes the Heisenberg time marking the onset of the SFF plateau. Furthermore, the late-time saturation value of complexity matches Hermitian model expectations~\cite{Erdmenger:2023wjg,Rabinovici:2020ryf,Rabinovici:2022beu}, with $C_{\sigma}(t=\infty) \approx d/2$.

\begin{figure}[t!]
 \centering
     {\includegraphics[width=4.2cm]{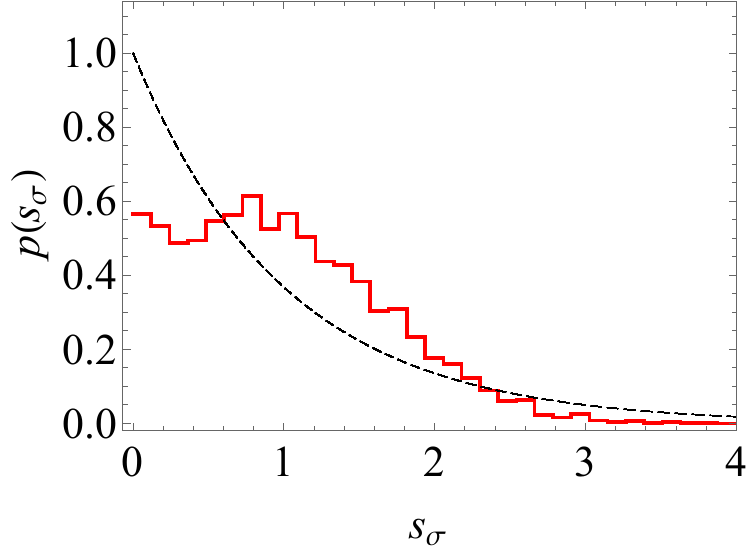}}
     {\includegraphics[width=4.2cm]{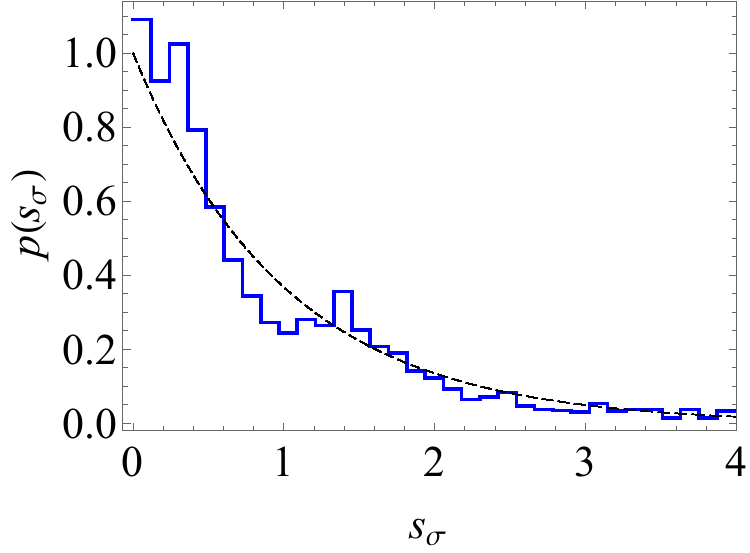}}
     
     {\includegraphics[width=4.2cm]{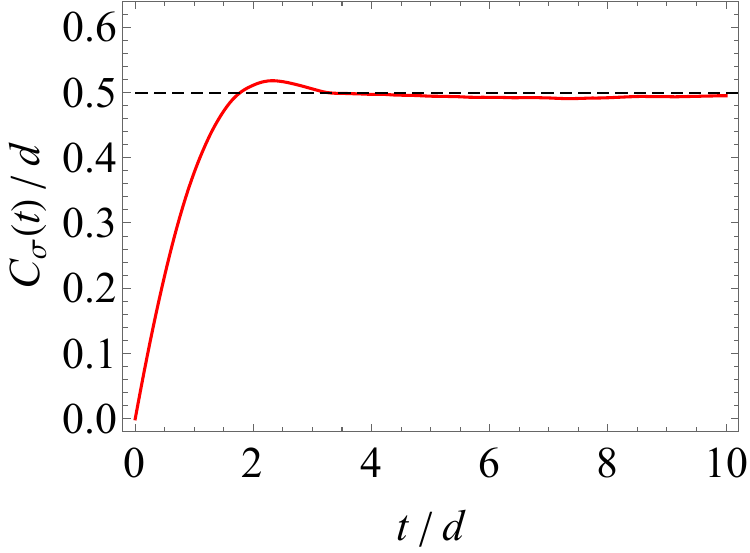}}
     {\includegraphics[width=4.2cm]{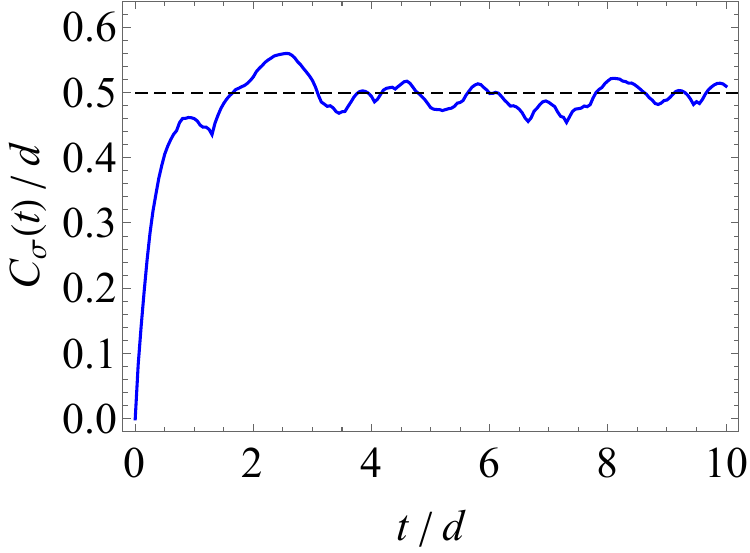}}
\caption{The top panels show the singular level spacing distribution for $q=4$ (red) and $q=2$ (blue) in the Hermitian limit of the model ($M\to0$), with dashed lines indicating the one-dimensional Poisson distribution. The bottom panels present the SVD Krylov complexity for both cases. Thus, while SVD methods fail to reproduce the eigenvalue-based results of the chaotic $q=4$ SYK model, they appear to correctly capture the behavior of the integrable $q=2$ case.} \label{SingularKrylovfig2}
\end{figure}

We note additional shortcomings of the SVD method. Specifically, quantities constructed using this approach fail to reproduce traditional eigenvalue-based results in the Hermitian limit ($M\to0$ in our model) for chaotic systems. For example, the $q=4$ model, known to be chaotic, exhibits a GUE distribution and a distinct peak in Krylov state complexity. However, as shown in Fig.~\ref{SingularKrylovfig2}, the singular value spacing distribution does not recover the expected GUE distribution. Instead, it appears neither Poissonian nor GUE-like, aligning with the Wigner surmise of singular value statistics~\cite{Kawabata:2023aa}. Moreover, no clear peak in Krylov complexity is observed.

The above behavior can be understood by noting that, in the Hermitian limit, singular values correspond to the absolute values of the eigenvalues. When the energy spectrum is not positive definite, as in the Hermitian SYK model, singular values may deviate from energy eigenvalues, leading to entirely different and erroneous conclusions. A similar argument applies to the $\langle r \rangle$ values in sparse non-Hermitian SYK models~\cite{Nandy:2024aa}.

In contrast, the Hermitian $q=2$ model is integrable, as indicated by its one-dimensional Poisson distribution and the absence of a peak in Krylov complexity~\cite{Baggioli:2024wbz}. Here, the SVD method also yields a Poissonian distribution, seemingly aligning with standard eigenvalue-based statistics. While this suggests that singular value statistics may diagnose integrability in Hermitian systems, its generality for systems exhibiting level clustering remains unclear, warranting further investigation.

%
\vspace{1mm}
\textbf{Conclusions.} 
By analyzing the prototypical non-Hermitian $q$-body SYK model, we identified key limitations of singular value decomposition methods in probing quantum chaos. Specifically, the SVD approach (I) fails to distinguish between chaotic and integrable non-Hermitian systems and (II) does not, in general, recover conventional eigenvalue statistics in the Hermitian limit when the spectrum is not positive definite. These findings suggest that the SVD methods are inadequate for characterizing quantum chaos or its lack thereof in general non-Hermitian systems. Our conclusion aligns with recent independent results on non-Hermitian random matrices~\cite{recent}, which appeared simultaneously with our work.

One might consider addressing the issue (II) above by rigidly shifting the eigenvalue spectrum to make it positive definite. In the Hermitian case, where the eigenvalues are real, this approach is indeed effective: applying the SVD method to the shifted spectrum yields the same Krylov complexity as the traditional EVD method. However, in the non-Hermitian case, this strategy fails, as the resulting Krylov complexity becomes highly sensitive to the arbitrary choice of the shift in the complex plane.

Our observations underscore the need for more robust and reliable methods to study quantum chaos in non-Hermitian systems, such as the bi-Lanczos approach for Krylov complexity. Recent studies in this context have focused on operator growth~\cite{Nandy:2024htc}, while others have examined state complexity in certain non-Hermitian models~\cite{Bhattacharya:2023yec,Bhattacharyya:2023grv}. Notably, Krylov state complexity, constructed via the bi-Lanczos approach, is shown to effectively diagnose quantum many-body chaos in a random non-Hermitian XY model~\cite{Zhou:2025ozx}.
Applying this approach to Krylov complexity in nHSYK could yield novel insights, particularly in holographic contexts, advancing our understanding of quantum chaos in non-Hermitian systems. We aim to explore this direction in the near future~\cite{newpaper_wip}.

%
\vspace{2mm}
\textit{Acknowledgments.}
We would like to thank {Johanna Erdmenger, Manas Kulkarni, Pratik Nandy and Yijia Zhou} for valuable discussions and correspondence.  
M. B. and K. B. H. acknowledge the support of the Foreign Young Scholars Research Fund Project (Grant No.22Z033100604). M. B. acknowledges the sponsorship from the Yangyang Development Fund.
H. S. J. and J. F. P. are supported by the Spanish MINECO ‘Centro de Excelencia Severo Ochoa' program under grant SEV-2012-0249, the Comunidad de Madrid ‘Atracci\'on de Talento’ program (ATCAM) grant 2020-T1/TIC-20495, the Spanish Research Agency via grants CEX2020-001007-S and PID2021-123017NB-I00, funded by MCIN/AEI/10.13039/501100011033, and ERDF `A way of making Europe.'
K. Y. K. was supported by the Basic Science Research Program through the National Research Foundation of Korea (NRF) funded by the Ministry of Science, ICT $\&$ Future Planning (NRF-2021R1A2C1006791) and the Al-based GIST Research Scientist Project grant funded by the GIST in 2025. K. Y. K. was also supported by the Creation of the Quantum Information Science R$\&$D Ecosystem (Grant No. 2022M3H3A106307411) through the National Research Foundation of Korea (NRF) funded by the Korean government (Ministry of Science and ICT). K. Y. K. was also supported by the Gwangju Institute of Science and Technology (GIST) research fund (Futureleading Specialized Resarch Project, 2025) and under the framework of international cooperation program managed by the National Research Foundation of Korea (RS-2023- NR119907, FY2023).

\vspace{2mm}
M. B., K.-B. H., H.-S. J, X. J., K.-Y. K, and J. F. P. contributed equally to this work.

\bibliographystyle{apsrev4-1}
\bibliography{Ref}
\end{document}